\documentclass[submission,copyright,creativecommons]{eptcs}
\providecommand{\event}{LAFM 2013} 
\usepackage{amsmath,amsthm,amssymb,graphicx,verbatim}
\usepackage{tikz} 
\usepackage{tikzYAWL} 
\usetikzlibrary{matrix,chains,decorations.text,positioning,shadows,arrows,shapes,shapes.geometric,petri,decorations.markings}

\newcommand{\TikzToolSymbol}[9]{%
\draw[rotate=-4,thick, drop shadow,fill=black!10] (0:#2)
\foreach \i in {1,...,#1} {%
  {[rotate=(\i-1)*360/#1] -- (0:#2)  arc (0:#4:#2) {
             -- (#4+#5:#3)  arc (#4+#5:360/#1-#5:#3)}} } -- cycle ;
\node [circle, thick, draw=black,fill=white,minimum width=.45cm] {}; 
\draw [-,draw=none,postaction={decorate,decoration={text along path, text={|\scriptsize \bf|#6}}}] (0:-#7) arc (0:-360:-#7);
\node [circle, minimum width=#9,draw=none,fill=none](#8){};
}

\newcommand{\TikzYAWLDocSymbol}[3]{%
\node (#3) [rectangle, drop shadow ,thick, draw=black,fill=white,minimum width=#1cm,minimum height=#2cm]{};
\node[ fill= black, text = white, right] at (#3.north west) {\scriptsize YAWL};
\node[ YAWLStartCondition, right=1pt] (start) at (#3.west){};
\node[ YAWLEndCondition, left=1pt] (end) at (#3.east){};
\node[ rectangle, draw](t1) at (0,-.2){};
\node[ rectangle, draw](t2) at (0,.2){};
\draw (start) -- (t1);\draw (start) -- (t2);
\draw (t1) -- (end);\draw (t2) -- (end);
}

\newcommand{\TikzFSPDocSymbol}[3]{%
\node (#3) [rectangle, drop shadow, thick, draw=black,fill=white,minimum width=#1cm,minimum height=#2cm,text width=#1cm]{\tiny{P= a- b- P. \\ $\parallel$S= q[1..3]:P.}};
\node[ fill= black, text = white, right] at (#3.north west) {\scriptsize FSP};
}

\newcommand{\TikzCounterExampleDocSymbol}[3]{%
\node (#3) [rectangle, thick, draw=black,fill=white,minimum width=#1cm,minimum height=#2cm,text width=#1cm]{\tiny{1- a; \hspace{.1cm}  3- b; \\  2- a;  \hspace{.1cm}  4-ER;}};
\node[ fill= black, text = white, right] at (#3.north west) {\scriptsize CounterEx};
\draw (#3.west) -- (#3.east);
\draw (#3.north) -- (#3.south);
}

\title{Fluent Logic Workflow Analyser:\\ A Tool for The Verification of Workflow Properties}
\author{Germ\'an Regis \qquad\qquad Fernando Villar \qquad\qquad Nicol\'as Ricci
\institute{ Departamento de Computaci\'on \\
            Facultad de Cs. Exactas Fco.-Qcas. y Naturales\\
            Universidad Nacional de R\'{\i}o Cuarto\\
            Argentina}
\email{\{gregis,fvillar,nricci\}@dc.exa.unrc.edu.ar}
}
\def\titlerunning{Fluent Logic Workflow Analyser}
\def\authorrunning{G.~Regis, F.~Villar, N.~Ricci}
\begin{document}
\maketitle

\begin{abstract}
In this paper we present the design and implementation, as well as a use case, of a tool for workflow analysis. The tool provides an assistant for the specification of properties of a workflow model. The specification language for property description is Fluent Linear Time Temporal Logic. Fluents provide an adequate flexibility for capturing properties of workflows. Both the model and the properties are encoded, in an automated way, as Labelled Transition Systems, and the analysis is reduced to model checking.
\end{abstract}

\section{Introduction}
The importance of efficiency in companies requires constant improvement to their organisational processes. This has led to the need for expressing such processes, typically referred to as \emph{workflows}, and to the proposal of various workflow languages. There exist many workflow languages, differing in their degree of formalisation (e.g., informal, only with a formal syntax, etc.), their corresponding approaches to workflow description (e.g., declarative or procedural), their expressivity (e.g., some support advanced conditional routing and some not), their support for automated analysis, etc. An aspect that we consider particularly important is formal semantics. This aspect is crucial for the analysis of models in the language, and is also strongly related to expressivity, since more expressive languages are more difficult to fully formalise. Furthermore, expressivity and automation in analysis are typically conflicting aspects, and the design of a good language involves the search of an adequate balance between these aspects. This applies not only to the language in which a workflow is expressed, but also to the language used for describing declarative properties of a workflow. The importance of declarative properties of workflows is acknowledged by several researchers (see for instance \cite{Karamanolis+2000,Pesic+2010,WongGibbons2011}). In particular, in \cite{Pesic+2010} a declarative approach to business process modelling and execution is proposed, where declarative behavioural properties of procedural workflow models are a central characteristic. 

In this paper, we present a tool for workflow analysis. This tool allows the user to describe properties over a workflow model and verify these properties in an automated way. The formal language used for property specification is a known temporal logic, \emph{fluent linear temporal logic} (FLTL) \cite{GiannakopoulouMagee2003}, which is well suited for formally expressing declarative properties of workflows \cite{Regis+2012}. 

Basically, FLTL provides a convenient way of expressing state properties of a labelled transition system, associated with the occurrence of events in the system. More precisely, FLTL extends LTL by incorporating the possibility of describing certain abstract states, called \emph{fluents}, characterised by events of the system. As defined in  \cite{MillerShanahan1999}, fluents are time-varying properties of the world, which hold at particular instants of time if they have been initiated by a triggering event (occurring at some earlier instant in time), and have not been terminated by any terminating event since their initiation. 

For the verification we employ Model Checking \cite{Clarke+2001bk}, a well established automated method for verifying properties of finite state systems. In order to apply this technique using the \emph{Labelled Transition System Analyser} (LTSA), our tool encodes workflow models as Labelled Transition Systems, following the characterisation presented in \cite{Regis+2012}. Given a property, the tool guarantees that it is satisfied, or generates violating workflow executions when the property does not hold, as is typical with model checking.

As the input language for workflow description, the tool adopts YAWL (Yet Another Workflow Language) \cite{Aalst+2005}. YAWL is a powerful workflow language based on the use of workflow patterns \cite{Aalst+2003}. It is considered an expressive formalism, as various works dealing with its expressivity in relation to other business process languages demonstrate \cite{Hofstede+2010bk}. Indeed, the use of YAWL allows us to ensure the usability of our tool for other workflow languages, in many cases via the use of available automated tools mapping other formalisms into YAWL. 

In the remainder of the paper we present the main features of the tool and exemplify their use for describing, specifying and analysing properties of workflow models. Then, we describe the tool as an aggregation of two modules: the encoding manager and the environment that assists in property specification and realises the integration between the encoder and the LTSA model checker. Finally we conclude with a discussion on our conclusions and future work.

\section{Tool usage}
Let us illustrate the use of the tool via a simple hypothetical workflow model. The model, as depicted in Fig.~1, describes the process of making a trip. This process begins with the registration task, then the customer can book a flight, hotel or car. When some (may be all) of them are booked with the corresponding task, the customer must pay for them. A simple property of this process may be that once some booking was made, then the payment must take place.

The use of the tool starts by opening the YAWL\footnote{YAWL is a free workflow modelling tool that can be downloaded from \texttt{http://sourceforge.net/projects/yawl/}} specification of the workflow. Our tool allows the user to import such a specification, showing it in a graphical way. Once a workflow description is opened, we can add intended properties, in our case the above mentioned one. The properties are formulas specified in FLTL. For the proposed property, one possible specification putting emphasis on the fluents usage, may be \texttt{[](someBook -> <>(pay.start))}. The formula establishes that whenever a booking occurs, the payment must take place. 

Using \emph{drag and drop}, we can shape the structure of the formula by incorporating the desired operands from the operators bar. The operands can be events of the model, i.e., \emph{start} and \emph{end} task events, or \emph{fluents}. Fluents are binary variables whose values depend on two sets of events: activating and deactivating events. In our case, the operands of the formula are: the fluent \texttt{someBook}, that captures the occurrence of some booking, and the reference to the start of the payment process through the \texttt{pay.start} event.  

In order to specify fluents, we use the fluent definition feature of the tool, starting with the new fluent definition (\emph{main menu} option \texttt{add fluent}). Then using the fluent activating or deactivating tools, we set the corresponding model events for each fluent. In case of \texttt{someBook}, we select the events that enable this fluent, i.e., the \texttt{start} events of the \texttt{flight}, \texttt{hotel} and \texttt{car} booking processes. Note that the \emph{start} events are depicted at the left of tasks and the \emph{end} at the right of them, i.e., work flows from left to right. In a similar way, using the deactivating tool, we can set up the events that turn off the selected fluent.

To assist us, the tool provides an auto-complete feature. This feature shows, when we write an operand of a formula, a pop up list containing possible events of the model. The list starts showing the names of tasks or conditions and then a choice for each event about them. Similar to the operators, the fluents can be incorporated to the property specification by means of drag an drop from the fluents list to the desired place in the formula. We can of course avoid these assistants and simply type the formula.

Another feature that the tool supports is a property specification assistant, that provides a set of templates, as shown in Fig.~\ref{PropertiesPatternsScreenShot}. These templates allow the user to instantiate a generic property about the system. To use these templates, we can navigate over a list of properties, with each one containing its own description. When a property is selected, for each parameter (operand of the underlying formula) a box and button are displayed for assigning the event or fluent of the model desired.
Our sample property corresponds to a \emph{response} property that asserts that, given two activities $A$ and $B$, ``whenever $A$ is executed, then $B$ has to be eventually executed afterwards''. If we wish to use this template, instead of the handmade specification, we simply instantiate the template by assigning the fluent \texttt{someBooking} and the event \texttt{pay.start} as the $A$ and $B$ parameters, respectively. 

Finally by pressing the \texttt{check Property} button, we can verify if the property holds in our workflow model. In order to store the fluents and properties specifications for future use, we can save our current job as a file by using the corresponding menu option.
\vspace{-.3cm}
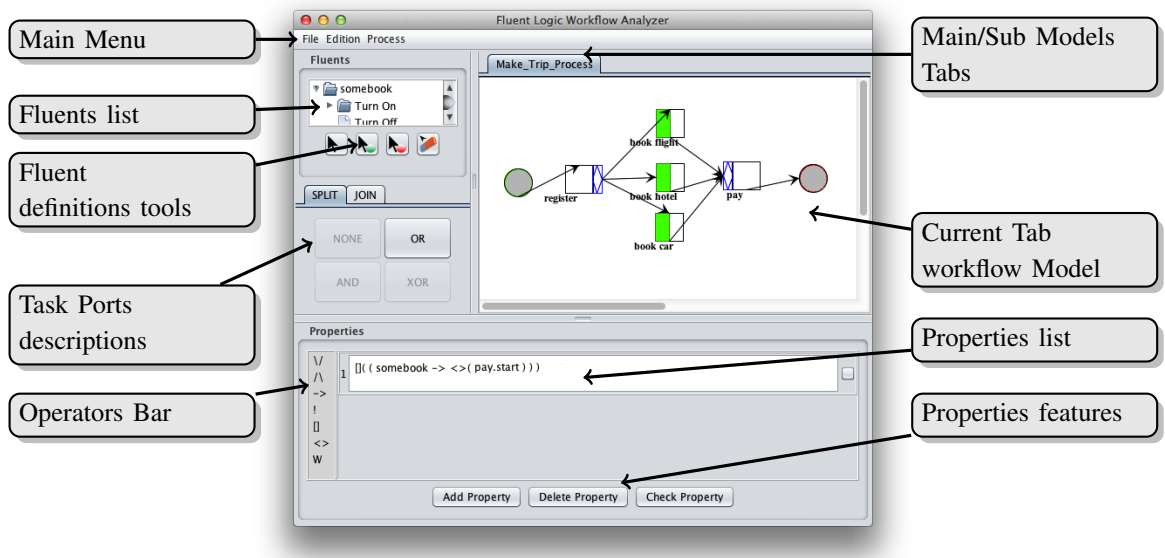
\begin{figure}[ht!]
\begin{center}
\tikzset{FloatingNodeText/.style={draw=black, fill= black!10, thick,rounded corners, text width=3cm, drop shadow}}
\tikzset{ArrowStyle/.style={->,very thick}}
\begin{tikzpicture} 
    \node[anchor=center,inner sep=0] at (0,-.3) {\includegraphics[width=9cm]{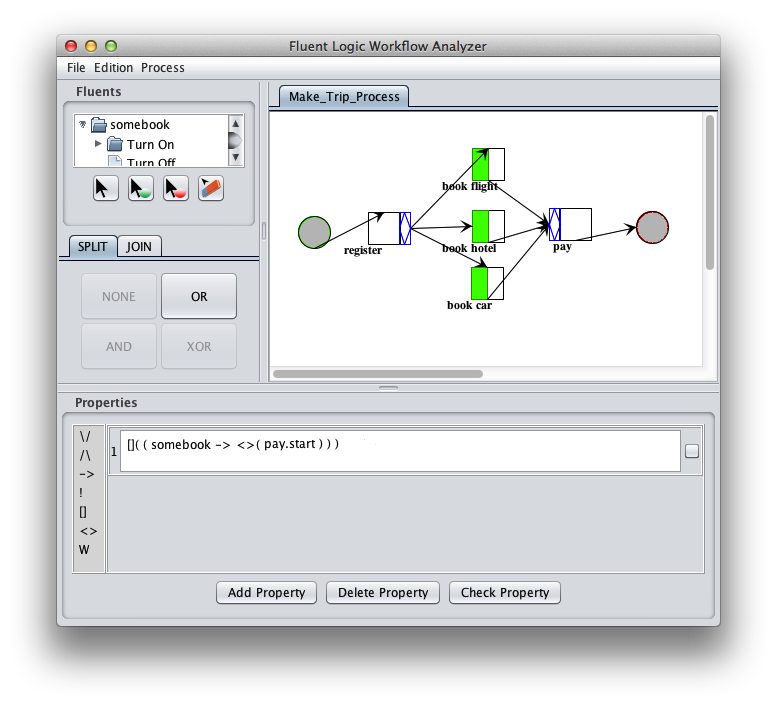}};
    \node[FloatingNodeText] (MainMenu) at (-6,3) {\small{Main Menu}};     
        \draw[ArrowStyle] (MainMenu) -- (-3.8,3);
    \node[FloatingNodeText] (FluentsList) at (-6,2) {\small{Fluents list}};   
        \draw[ArrowStyle] (FluentsList) -- (-3.5,2.1);
    \node[FloatingNodeText] (FluentsDefTools) at (-6,1) {\small{Fluent \\definitions tools}};
        \draw[ArrowStyle] (FluentsDefTools) -- (-3,1.6);
    \node[FloatingNodeText] (PortsDescription) at (-6,-0.8) {\small{Task Ports \\descriptions}};
        \draw[ArrowStyle] (PortsDescription) -- (-3.6,0.3);
    \node[FloatingNodeText] (OperatorsBar) at (-6,-2) {\small{Operators Bar}};
        \draw[ArrowStyle] (OperatorsBar) -- (-3.65,-1.6);
    \node[FloatingNodeText] (ProcessTabs) at (6,2.8) {\small{Main/Sub Models Tabs}};
        \draw[ArrowStyle] (ProcessTabs) -- ( 0,2.8);
    \node[FloatingNodeText] (CurrentTabModel) at (6,.2) {\small{Current Tab \\workflow Model}};
        \draw[ArrowStyle] (CurrentTabModel) -- ( 3,0.7);
    \node[FloatingNodeText] (PropertiesList) at (6,-1) {\small{Properties list}};
        \draw[ArrowStyle] (PropertiesList) -- ( 0,-1.5);
    \node[FloatingNodeText] (PropertiesFeatures) at (6,-2) {\small{Properties features}};
        \draw[ArrowStyle] (PropertiesFeatures) -- ( .5,-2.9);
\end{tikzpicture}
\vspace{-.4cm}
\label{FLYAnalyserScreenShot}
\caption{FLYAnalyser Full view}
\vspace{-.3cm}
\end{center}
\end{figure}

\begin{figure}[ht!]
\begin{center}
\tikzset{FloatingNodeText/.style={draw=black, fill= black!10, thick,rounded corners, text width=3cm, drop shadow}}
\tikzset{ArrowStyle/.style={->,very thick}}
\begin{tikzpicture} 
    \node[anchor=center,inner sep=0] at (0,0) {\includegraphics[width=10cm]{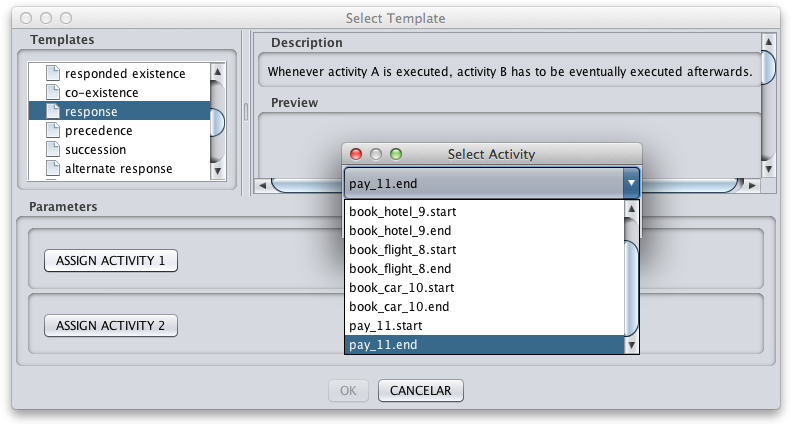}};
    \node[FloatingNodeText] (TemplatesList) at (-6.5,2) {\small{Properties \\Templates list}};     
        \draw[ArrowStyle] (TemplatesList) -- (-4.5,1.8);
    \node[FloatingNodeText] (ParametersList) at (-6.5,0) {\small{Template Parameters}};     
        \draw[ArrowStyle] (ParametersList) -- (-4.7,-1);
    \node[FloatingNodeText] (CurrentPropertyDescription) at (5,2.4) {\small{Current property \\description}};     
        \draw[ArrowStyle] (CurrentPropertyDescription) -- (0,2);
    \node[FloatingNodeText] (ParameterAssignment) at (5.1,0) {\small{Parameter Activity Assignment}};     
        \draw[ArrowStyle] (ParameterAssignment) -- (1.5,0.6);
\end{tikzpicture}
\vspace{-.2cm}
\label{PropertiesPatternsScreenShot}
\caption{Properties Template wizard}
\vspace{-.3cm}
\end{center}
\end{figure}
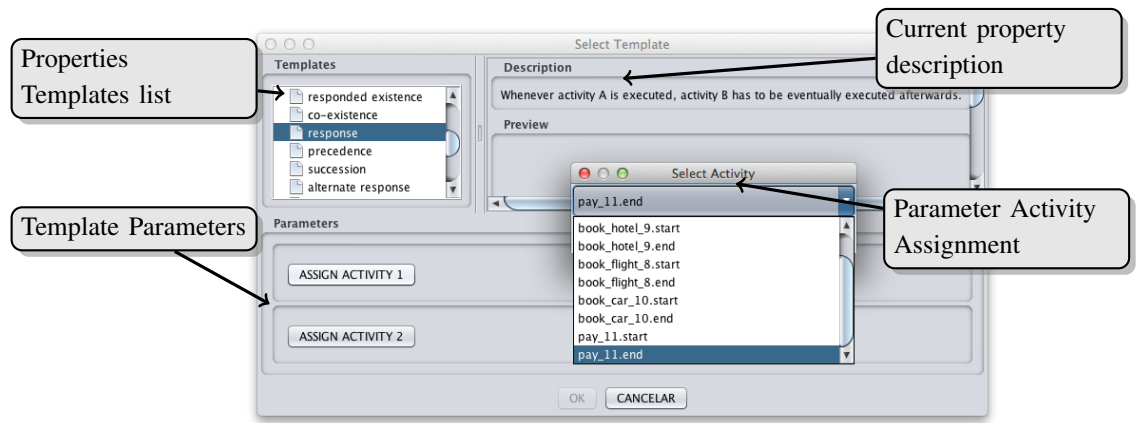

Note that the sets for activating or deactivating a fluent can be conformed by events corresponding to different sub-workflows, i.e., events of the workflow detailing a composite task. This flexibility allows the user to capture, in a simple way, complex situations in a model, such as for example execution traces between activities. 

\section{Tool Design}

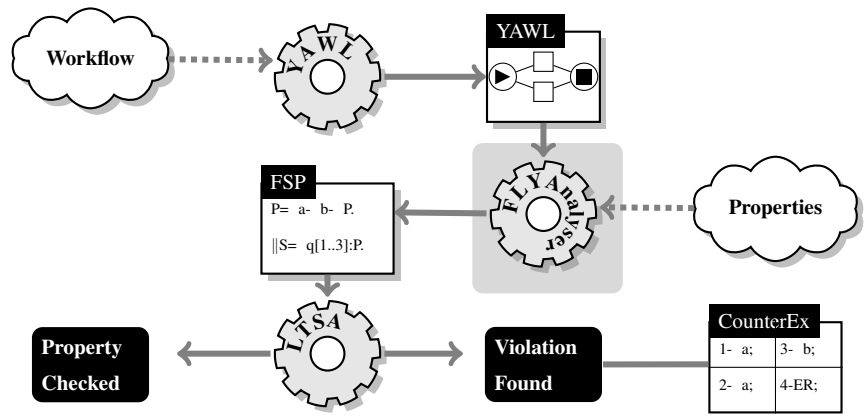
\begin{figure}[ht!]
\begin{center}
\begin{tikzpicture}[node distance=.3cm, bend angle=45 , auto]

\node[rectangle,rounded corners, fill=black!15, minimum width= 2cm, minimum height=2cm] (Resultado) at (1.5,-0.1){};


\matrix (Mat) [matrix of nodes, row sep=.3cm, column sep = 1.2cm] {  
\node [fill=white,thick,drop shadow, draw, cloud, ,cloud puffs=10,cloud puff arc=120,aspect=2, minimum height = .8cm, minimum width = 2cm] (ProcesoN) at (0,.15){\scriptsize \bf Workflow}; &
  \TikzToolSymbol{10}{.7}{.6}{20}{2}{YAWL}{.35}{YawlTool1}{1.5cm}; &
  \TikzYAWLDocSymbol{1.5}{1.2}{YawlDoc1}; & \\
   & \TikzFSPDocSymbol{1.5}{1.2}{FspDoc1}; & \TikzToolSymbol{10}{.7}{.6}{20}{2}{FLYAnalyser}{.35}{FLYATool1}{1.5cm}; & \node [fill=white,thick,drop shadow, draw, cloud, ,cloud puffs=10,cloud puff arc=120,aspect=2, minimum height = .8cm, minimum width = 2cm] (Reglas) {\scriptsize \bf Properties};\\ 
  \node[thick, rounded corners, fill=black, text=white, minimum width = 1.2cm, minimum height = 1cm,text width = 1.3cm](si) {\scriptsize \bf Property Checked}; & \TikzToolSymbol{10}{.7}{.6}{20}{2}{LTSA}{.35}{LtsaTool}{1.5cm}; & \node[thick, rounded corners, fill=black, text=white, minimum width = 1.2cm, minimum height = 1cm,text width = 1.3cm](no) {\scriptsize \bf Violation Found};& \TikzCounterExampleDocSymbol{1.5}{1.2}{CExDoc1};  \\ 
};     
\draw [-to, line width=0.08cm,dotted, draw=black!50] (ProcesoN) [yshift=.7cm]-- (YawlTool1);
\draw [-to, line width=0.08cm,dotted, draw=black!50] ( Reglas) [yshift=.3cm]-- (FLYATool1); 
\draw [-to, line width=0.08cm, draw=black!50] (YawlTool1) -- (YawlDoc1);
\draw [-to, line width=0.08cm, draw=black!50] (YawlDoc1.south) -- (FLYATool1) ;
\draw [-to, line width=0.08cm, draw=black!50](FLYATool1) [yshift=.5cm]-- (FspDoc1) ;
\draw [-to, line width=0.08cm, draw=black!50] (FspDoc1) --  (LtsaTool);
\draw [-to, line width=0.08cm, draw=black!50] (LtsaTool)--  ++(-2,0);
\draw [-to, line width=0.08cm, draw=black!50] (LtsaTool)--  ++(1.8,0);
\draw [line width=0.08cm, draw=black!50] (no)[yshift=.1cm] -- (CExDoc1);
\end{tikzpicture}
\vspace{-.3cm}
\caption{Verification Workflow}
\vspace{-.3cm}
\label{FBPVerificationMainDiagram-Figure}
\end{center}
\end{figure}

Fig.~\ref{FBPVerificationMainDiagram-Figure} depicts the architectural design of the process of verifying properties about a workflow model. The process begins with a model of workflow described using the YAWL workflow modelling tool. YAWL is a powerful workflow language based on the use of workflow patterns \cite{Aalst+2003}. In order to verify properties of workflows, in particular those described with YAWL, we develop a tool called FLYAnalyser. Our tool takes as input a workflow model and assists the user to easily and in a graphical way, specify and verify properties of the model. The verification process is handled by the LTSA model checker, through a translation of the workflow and properties to a labelled transition system (LTS). Thus we will be able to express behavioural properties of these workflows declaratively, using the FLTL. 

The tool has two main modules developed separately: a compiler called YAWL2FSP that encodes a YAWL workflow model into an LTS and the environment FLYAnalyser, which is responsible of the analysis and specification of properties, and handles the integration between the YAWL2FSP and LTSA.

\subsection{YAWL2FSP}
As a module of the FLYAnalyser, we first present an encoding from YAWL models into Finite State Processes (FSP)\cite{MageeKramer2006bk}. FSP is a process algebra that provides us a compact and flexible way to specify LTS, i.e., FSP expressions can be automatically mapped into LTS. Basically, this encoding allows us to interpret YAWL procedural workflows as FSP processes. The encoding will enable us to employ the LTSA model checker for \emph{verifying} behavioural properties of task activities of the workflow specifications. The basic intuition behind the encoding of a YAWL models into FSP is the following. A system's behaviour is characterised by the occurrence of its tasks. In an abstract way, we can capture a task as an entity having some activity in the system between its \emph{start} and \emph{end} events. So, a trace of these events describes a possible execution of the system. In this way, a system's behaviour is captured by the set of all its execution traces. These traces are obviously constrained according to the control flow of the system. 

According to our previous observation, it is straightforward to see that a task activity can be captured by means of a \emph{fluent}, becoming \emph{true} when its \emph{start} event takes place, and turning back to \emph{false} when its \emph{end} event task occurs. In order to capture the behaviour of the workflow's control flow, we will need to introduce appropriate event synchronisations and process compositions, relating the events related to the tasks that conform the workflow. Once we achieve a characterisation of workflows as FSP processes, we can express properties of the workflows by expressing temporal formulas, employing task-related fluents as the basic ingredient. 

To formally describe our translation from YAWL into FSP, we consider a formal semantics of YAWL models \cite{Hofstede+2010bk}, given in terms of Reset Petri Nets. Taking into account this semantics, we propose an encoding for tasks and conditions. In order to represent a workflow behaviour, we specify how to compose tasks and conditions. In this composition we consider the control flow operators associated with the tasks of the workflow, and provide an encoding for them. Finally, we address especially sophisticated elements of YAWL constructions, such as \emph{cancel regions}, \emph{multiple instance tasks} and \emph{composite tasks}. For a full description of the encoding process, see \cite{Regis+2012}.

\subsection{The Environment}

The environment is a graphical application written in Java. It can open a workflow model specified with the YAWL tool. These models are saved in files with XML format. 
The application has the following features:
\begin{itemize}
  \item It shows the model in a graphical way to help the user understand and analyse the model. In presence of composite task sub-workflows specifications, for each of them, the application generates a tab with the corresponding graphical workflow specification.
  \item It provides a flexible and agile way to specify fluents by simply clicking on the desired events of the workflow that activate or deactivate them. 
  \item It assists the user in the property specification by means of \emph{drag and drop} operators on fluents, and auto-completing with events of the models.
  \item It aids the workflow analysis by means of a property template wizard.
  \item It holds the integration with other modules and tools, in particular, it invokes the compiler to obtain the FSP model of the workflow and composes it with the selected properties. Then, it calls the model checker in order to perform the verification.  
\end{itemize}

Note that the property templates listed by the wizard are those analysed in \cite{Pesic+2010} and they are stored in the file \texttt{templates.properties} in XML format. The user can incorporate new templates by adding them into the file, filling certain required fields. 

\section{Conclusion and Future Work}
We have described the \emph{Fluent Logic Workflow Analyser}, a tool for the specification and verification of workflows and properties of these. The tool focuses on the analysis of \emph{declarative properties} of procedural descriptions of workflows. This tool is publicly available\footnote{\texttt{http://sourceforge.net/projects/yawl2fsp/}}.

We have chosen to base our work on YAWL because it has a formal foundation, and it supports a wide range of workflow patterns, providing an expressive environment for workflows specification. The YAWL toolset provides the verification of some properties of workflows such as soundness and deadlock-freedom \cite{Aalst+2011}, but it does not provide a suitable flexible language for declaratively expressing other behavioural properties of its models. Our tool complements YAWL toolset in this direction.

In order to analyse the scalability of our tool, obviously bounded by the state explosion problem of the underlying model checking technique, we took as a case study a relatively complex example from \cite{Hofstede+2010bk}. The case study describes the process of order fulfilment which is divided into several phases. The complete model, flatting the composed tasks, has 58 tasks, 30 gates, 36 conditions, 2 cancel regions. The LTS was generated in 0.298 seconds, using 28,96 Mbytes of memory, containing 13164 states and 59722 transitions. Several properties were verified and the time consumption associated with this process did not exceed one second for each one. Due to space restrictions, we do not describe this model here. 

As a future extension of the tool, we are working on the feedback of the information provided by the model checker in case of property violations, showing a graphical representation of the sequence of events that produces a property violation. We think that this feature can help the user to understand why the model is wrong with respect to some analysed property.  

\bibliographystyle{eptcs}
\bibliography{Bibliografia}
\end{document}